\documentclass[10pt,twocolumn,letterpaper]{article}

\usepackage[pagenumbers]{wacv} % To force page numbers, e.g. for an arXiv version

\usepackage{newtxtext}
\definecolor{wacvblue}{rgb}{0.21,0.49,0.74}
\usepackage[pagebackref,breaklinks,colorlinks,allcolors=wacvblue]{hyperref}
\usepackage[most]{tcolorbox}
\usepackage{makecell}
\usepackage{hyperref}
\usepackage{url}
\usepackage{bbm}
\usepackage{caption}
\usepackage{quickmath}
\usepackage{lipsum}
\usepackage[ruled,vlined]{algorithm2e}
\SetKwComment{Comment}{/*}{*/}
\usepackage{array}
\usepackage{soul}
\usepackage{titletoc}
\usepackage{fontawesome5}

\usepackage[table]{xcolor}
\definecolor{lightpurple}{RGB}{180,200,230}
\definecolor{linkblue}{RGB}{70,130,180}
\usepackage{enumitem,amssymb}
\usepackage{wrapfig}
\usepackage{multirow}  % For multirow support
\usepackage{graphicx}
\newlist{todolist}{itemize}{2}
\setlist[todolist]{label=$\square$}
\usepackage{pifont}
\usepackage{multirow}
\usepackage{tabularx,threeparttable,array}
\usepackage{algorithmic}
\usepackage{xspace}

\newcolumntype{P}{>{\centering\arraybackslash}X}

\newcolumntype{Y}{>{\centering\arraybackslash}X}

\newcolumntype{Z}{>{\columncolor{lightpurple!25}\centering\arraybackslash}X}

\newcommand{\ours}{\textsc{PALM}\xspace}

\def\wacvPaperID{995} % *** Enter the WACV Paper ID here
\def\confName{WACV}
\def\confYear{2027}

\title{Learning to See Locally and Align Clinically with Pathology Semantics for Radiology Report Generation}

\author{Xuan Cuong Ngo \\
{\tt\small ngoquy12332115@gmail.com}}
\begin{document}
\maketitle
\begin{abstract}
Recent radiology-adapted vision-language models have achieved strong performance on standard report generation benchmarks, yet their robustness and generalization remain constrained by imperfect alignment and correlation between visual and textual features. Existing methods connect image and text either implicitly through autoregressive report supervision or explicitly through contrastive learning. However, autoregressive supervision alone is insufficient to establish reliable image-text alignment, while contrastive learning can push apart unpaired reports that describe related pathologies because they are not paired with the same image. This is problematic in radiology, where different reports may describe compatible pathology semantics even when they are unpaired. Misalignment between image and report representations can prevent the model from capturing key radiographic evidence. As a result, the decoder receives insufficient visual cues and tends to fill in the missing information using pretrained language priors, generating clinically plausible but visually unsupported reports.
% As a result, the learned representation may fail to organize images and reports around shared pathology concepts \hl{organize images and ...?}, causing the decoder to rely on pretrained language priors and generate clinically plausible reports that are not fully supported by radiographic evidence  \hl{this sentence need to be rewritten}.
To address this issue, we propose \ours, a pathology-aware alignment framework for radiology report generation. Instead of directly matching each image-report pair while separating all others, \ours aligns visual and textual features through shared pathology prototypes. These prototypes provide a clinically meaningful bridge between radiographic evidence and textual findings, allowing cases with similar pathology semantics to move toward common concepts without separating compatible cases. In addition, we introduce Masked Evidence Modeling to strengthen the image encoder’s sensitivity to local radiographic evidence by learning semantic changes caused by masked image regions. Experiments on MIMIC-CXR, IU X-Ray, and MIMIC-ABN show that \ours consistently improves both report generation and abnormality-focused robustness. The code and models will be made publicly available.

% \hl{Code and models will be publicly available}.
\end{abstract}
    
\section{Introduction}

Chest X-ray (CXR) is one of the most utilized imaging techniques in modern healthcare~\cite{irvin2019chexpert,sei}. Given a CXR, radiologists examine each depicted anatomical region and describe findings in a detailed textual report. Given the large volume of CXRs to be examined in daily clinical practice, this often becomes a time-consuming task, which is further exacerbated by a shortage of trained radiologists in many healthcare systems~\cite{rosenkrantz2016us, rimmer2017radiologist, bastawrous2017improving}. As a result, \textit{radiology report generation} (RRG)~\cite{raoof2012interpretation, johnson2019mimicphysio} has emerged as an active research area to automatically analyze medical images and alleviate radiologists' workload. 

Recent medical vision-language models (MVLMs)~\cite{liu2023visual, li2023llava, zambrano2025clinically} have achieved impressive benchmark performance on radiology report generation (RRG) tasks. However, many recent efforts improve RRG mainly by scaling model size or incorporating additional information, such as retrieval-augmented generation (RAG) \cite{park2025dart} or multi-view inputs \cite{liu2025enhanced}. Despite these advances, current models still tend to rely heavily on language priors rather than visual evidence \cite{srivastava2026cwcd}. When visual evidence is weakly encoded or insufficiently aligned with the language decoder, generated reports become more dependent on learned language regularities than on image-specific findings. This can lead to clinically plausible but visually unsupported descriptions, a limitation observed in hallucination and factual-consistency studies of radiology report generation and vision-language models \cite{miura2021improving,ramesh2022improving,zhang2411radflag,srivastava2026cwcd}.

% To investigate how visual information influences MVLM responses in RRG, we analyze model behavior under visual information absence. As shown in Fig.~\ref{fig:teaser}, the model produces highly similar reports for both the original image and the occluded image, even when the most clinically relevant region in the second image has been masked out. This observation suggests that current RRG models may not sufficiently rely on radiographic evidence when generating reports. 

\begin{figure}[t]
    \centering
    \includegraphics[width=\linewidth]{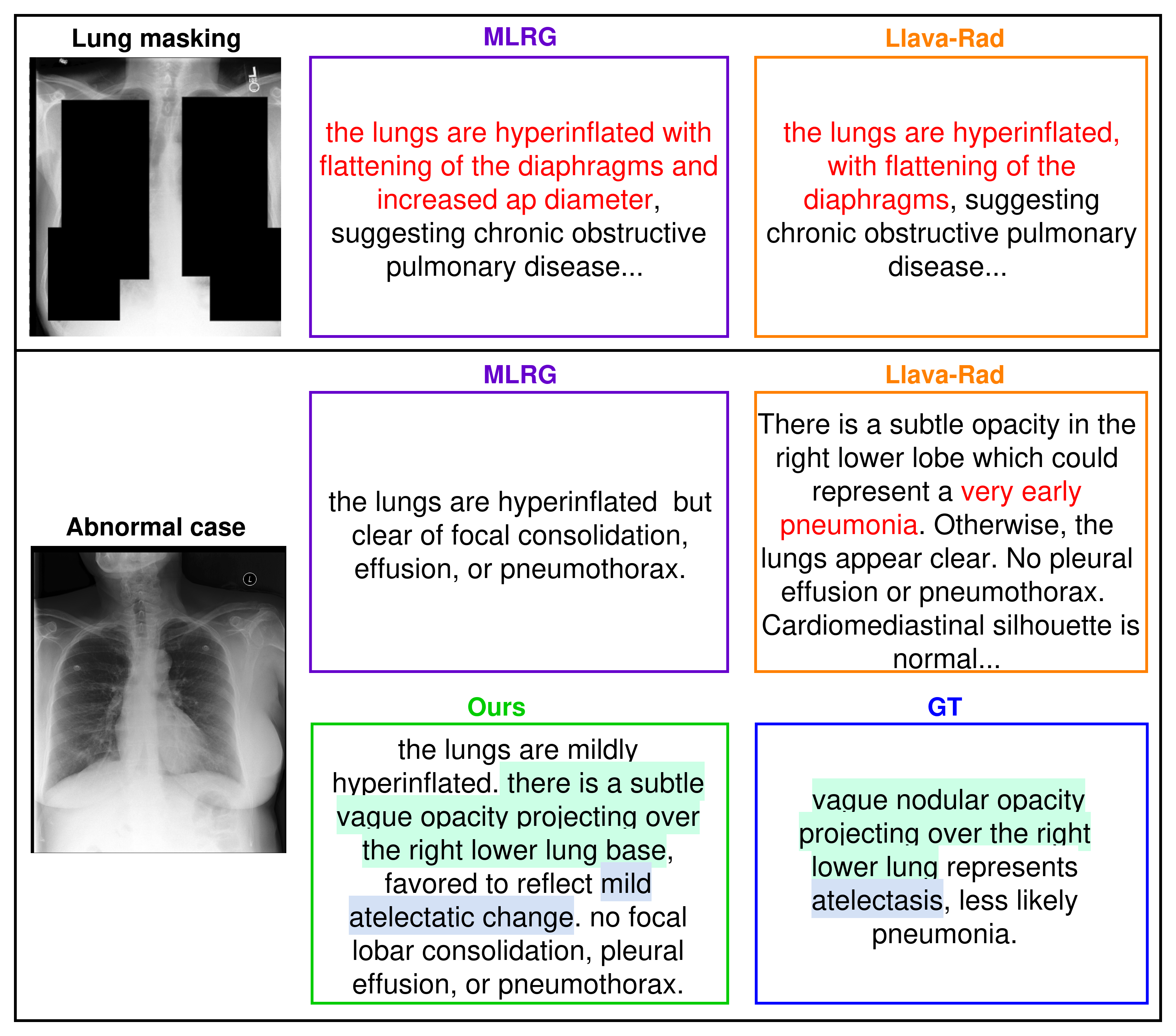}
    \caption{\textbf{Model responses under visual information loss and abnormal findings.} When lung regions are masked, existing MVLM-based RRG models still describe lung-related findings despite the removal of visual evidence. On an abnormal case, they produce plausible but incomplete reports that miss the true finding, whereas our report is more consistent with the ground truth.}
    \label{fig:teaser}
    \vspace{-1.5em}
\end{figure}

To investigate how visual evidence affects MVLM-based RRG, we study model behavior under regional evidence masking. As shown in Fig.~\ref{fig:teaser}, when lung evidence is partially occluded, the models still generate detailed statements about the masked regions. This suggests that the generated reports are often driven by learned reporting patterns from the training corpus, instead of sufficiently attending to the visible radiographic clues. Consequently, for abnormal cases, the models tend to produce plausible and commonly occurring descriptions while failing to recognize the actual abnormal finding, leading to reports that miss the clinically relevant evidence.

We argue that this limitation is partly caused by insufficient alignment between visual and textual representations in current training paradigms. Existing RRG models commonly establish image-text correspondence through autoregressive report supervision and, in some cases, contrastive learning. Autoregressive supervision optimizes the generated report sequence, but it does not explicitly ensure that each clinical statement is supported by the corresponding image evidence. Contrastive learning can strengthen image-text matching by pulling paired image-report representations closer. However, its negative-pair objective may also push apart unpaired samples within a batch, even when their reports describe related or partially overlapping pathologies. 
This assumption is problematic in radiology because unpaired reports are not necessarily clinically dissimilar. Different patients may share the same radiographic findings, such as pleural effusion or cardiomegaly, even though their reports are not paired with the same image. This inherent behavior of contrastive learning can lead to a modality gap between visual and textual representations, consistent with the observation of Chowers et al.~\cite{chowers2026modality}.

Consequently, current RRG models may learn an incomplete multimodal alignment, causing the decoder to rely disproportionately on language priors while overlooking important visual information. This weak visual dependence reduces clinical specificity and increases the risk of hallucinated or unsupported findings. Therefore, improving pathology-aware image-text alignment is essential for generating radiology reports that are both fluent and faithful to the input images.

To address this problem, we propose \ours{}, a pathology-aware alignment framework for RRG. \ours{} contains two complementary modules: a prototype-based alignment module that aligns image and text representations at the global pathology-semantics level, and a Masked Evidence Modeling (MEM) module that strengthens fine-grained visual evidence modeling. Instead of directly aligning each image--report pair while treating all other samples as negatives, the prototype-based alignment module introduces pathology-aware prototypes as shared semantic anchors across the visual and textual modalities. By aligning both image and report representations with these prototypes, \ours{} allows cases with similar pathology semantics to share common clinical concepts without pushing compatible samples apart. This reduces the modality gap while preserving pathology structures shared across different cases.

Complementary to this global alignment, \textbf{Masked Evidence Modeling (MEM)} enhances the visual encoder’s sensitivity to local radiographic evidence. Here, local radiographic evidence means that each visual patch should encode the clinical cues present in its corresponding image region, such as localized opacity, boundary changes, texture patterns, or lesion-related appearance that may support a radiographic finding. MEM encourages each visual patch token to encode the semantic contribution of its corresponding image region by learning how masked regions affect the image-level semantic representation. As a result, the model learns to connect fine-grained radiographic details with pathology-level concepts, supporting more faithful and clinically informative report generation.

Our contributions are summarized as follows:
\begin{itemize}
\item We identify the modality misalignment problem in RRG and propose \ours{}, a pathology-aware alignment framework that aligns image and text representations through shared pathology prototypes, enabling semantically related cases to be structured by common clinical concepts rather than being incorrectly pushed apart.

\item We introduce MEM , a self-distillation objective that encourages patch-level visual tokens to encode the semantic contribution of local image regions, strengthening the model's use of fine-grained radiographic evidence during report generation.

\item We conduct extensive experiments on MIMIC-CXR, MIMIC-ABN, and IU X-Ray, demonstrating that \ours{} achieves state-of-the-art performance and generates radiology reports that are more faithful to the underlying visual evidence.
\end{itemize}

\section{Related Works}

\noindent
\textbf{Radiology Report Generation.}
Radiology report generation (RRG) has evolved from CNN-RNN architectures~\cite{wang2018tienet, jing2018automatic, li2018hybrid, jing2019show, zhang2020radiology, xue2018multimodal, yin2019automatic} adapted from image captioning~\cite{xu2015show, vinyals2015show, you2016image, cornia2020meshed} to Transformer-based frameworks~\cite{chen2020generating, liu2021exploring, you2021aligntransformer, liu2021contrastive, chen2021cross, vaswani2017attention}. Recent methods improve report generation using memory mechanisms~\cite{chen2020generating, chen2021cross, shen2024automatic}, contrastive learning~\cite{liu2021contrastive, li2023dynamic}, knowledge graphs~\cite{liu2021exploring, zhang2020radiology, wang2022medical}, region guidance~\cite{tanida2023interactive}, disease tags~\cite{wang2022medical, you2021aligntransformer}, pretrained models~\cite{nicolson2022improving}, large language models~\cite{liu2024bootstrapping, liu2024context}, and multi-view or longitudinal information~\cite{nicolson2023-longitudinal-multiview, tmm_mulview_2024-fmvp, 2024-eccv-hergen}. While these approaches improve generation quality and clinical accuracy, they mainly rely on report-level supervision or paired image-text matching. In contrast, our work introduces pathology-aware visual-textual alignment, where image and text representations are structured by shared clinical prototypes rather than only paired-sample supervision.

\noindent
\textbf{Vision-Language Models in Medical Imaging.}
Large multimodal models~\cite{liu2023visual, alayrac2022flamingo} connect visual encoders with language models for general vision-language instruction following. In medical imaging, vision-language pretraining methods~\cite{wang-mgca, zhang-kad, Wu-medclip} and medical adaptations such as LLaVA-Med~\cite{li2023llava} and LLaVA-Rad~\cite{zambrano2025clinically} extend this paradigm to clinical tasks, including radiology report generation. However, their visual representations are often learned through global image-text objectives, providing limited supervision for local radiographic evidence. Our work addresses this issue through pathology-aware alignment and Masked Evidence Modeling, which strengthen semantic alignment and fine-grained visual representation learning.

% \noindent
% \textbf{Vision-Language Models in Medical Imaging.}
% Large multimodal models~\cite{liu2023visual, alayrac2022flamingo} connect pretrained visual encoders with language models through projection modules, enabling general vision-language instruction following. In medical imaging, vision-language pretraining methods~\cite{wang-mgca, zhang-kad, Wu-medclip} learn transferable representations through multi-granularity alignment, knowledge integration, or semantic matching, while medical adaptations such as LLaVA-Med~\cite{li2023llava} and LLaVA-Rad~\cite{zambrano2025clinically} extend this paradigm to clinical tasks including radiology report generation. Although these models provide strong language generation ability, their visual representations are often optimized through global image-text objectives, which may provide limited supervision for local radiographic evidence. Our work addresses this limitation from the perspective of pathology-aware visual-textual alignment, combining shared pathology prototypes for semantic organization with Masked Evidence Modeling to strengthen fine-grained visual representation learning.

\section{Methodology}
\begin{figure*}[t]
    \centering
    \includegraphics[width=\linewidth]{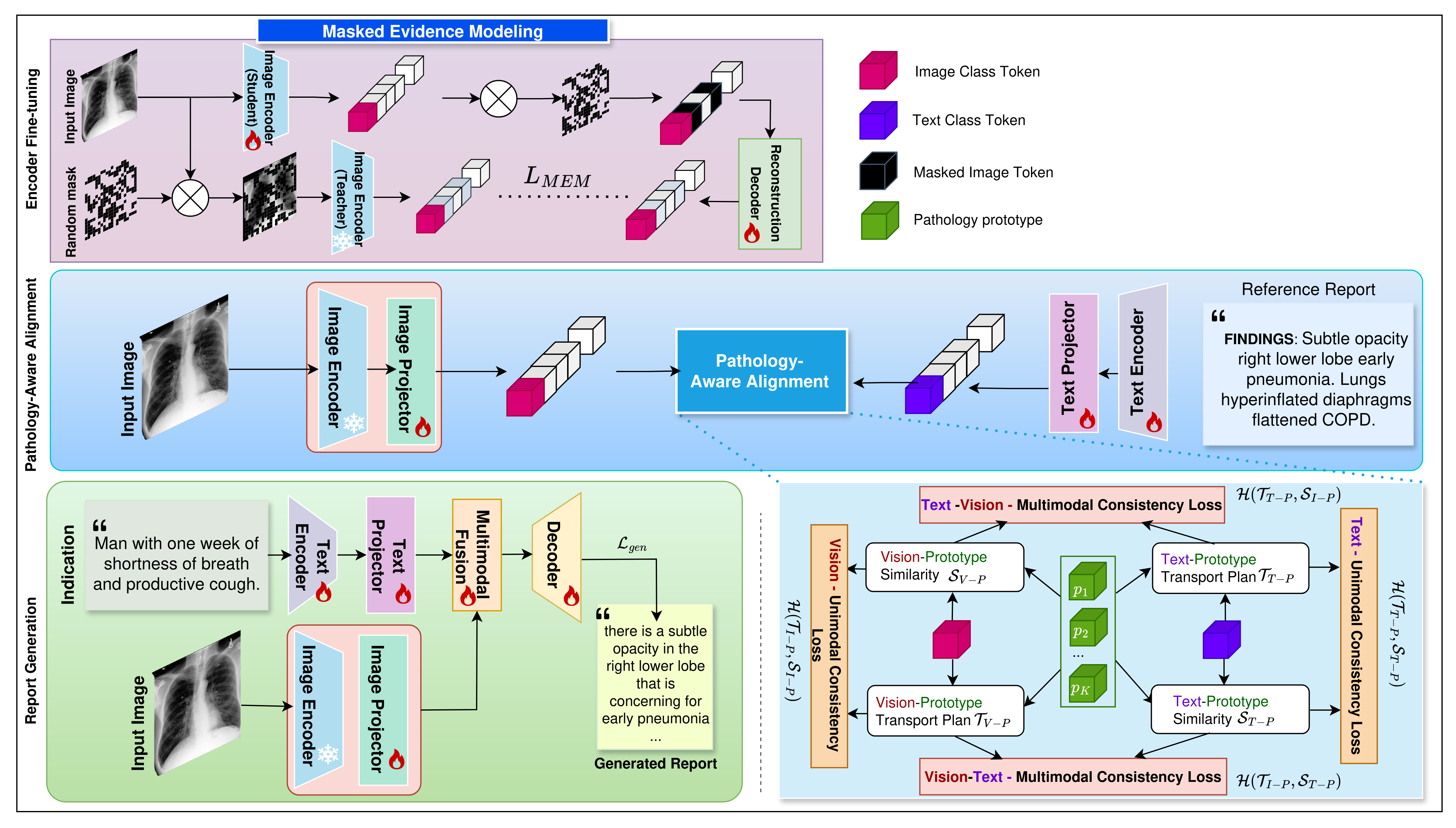}
    \caption{
    Overview of \ours{}. The framework first fine-tunes the visual encoder with Masked Evidence Modeling, which encourages image tokens to encode local radiographic evidence under random masking. It then performs multimodal alignment through the proposed Pathology-Aware Alignment module, where visual and textual representations are aligned via shared pathology prototypes. The learned features are finally used, together with the clinical indication, to generate the radiology report.
    }
    \label{fig:main}
\end{figure*}
Figure~\ref{fig:main} presents an overview of our proposed \textbf{\ours} framework. \ours{} contains three components: \textbf{Masked Evidence Modeling (MEM)}, \textbf{Pathology-Aware Alignment}, and autoregressive report generation. \ours{} first fine-tunes the visual encoder with \textbf{Masked Evidence Modeling (MEM)} to enhance its sensitivity to local radiographic evidence. It then performs \textbf{Pathology-Aware Alignment}, which bridges image and text representations through shared pathology prototypes, allowing semantically related cases to move toward common clinical concepts without forcing compatible samples apart. Finally, the aligned multimodal representations are optimized with a standard autoregressive objective for RRG.

\subsection{Masked Evidence Modeling}
RRG requires the model to understand not only the global content of an image, but also the local visual evidence that supports each clinical finding. However, a vision encoder trained mainly with image-level supervision may produce strong global representations, while its patch-level tokens may fail to capture strong local features. In this case, the language model may receive visual tokens that are not sufficiently informative about local radiographic details. To address this issue, we introduce \textbf{Masked Evidence Modeling (MEM)}, a visual representation learning objective that encourages each patch token to encode the semantic contribution of its corresponding image region.

The intuition of MEM is simple. If a clinically important region is removed from the image, the semantic representation of the image should change. For example, if a lung opacity region is masked, the image representation should no longer contain the same evidence for opacity. Therefore, by repeatedly masking different regions and observing how the semantic representation changes, the model can learn what information each local region contributes. 
% Unlike MAE \cite{he2022masked} methods that reconstruct missing pixels, MEM reconstructs the semantic embedding produced by a frozen teacher encoder under masked visual input. This encourages the model to learn stronger local features instead of relying mainly on global image features.

Specifically, MEM uses two copies of the visual encoder: a frozen teacher encoder and a trainable student encoder. The teacher encoder is initialized from the pretrained visual backbone and kept fixed during training. Its role is to provide persistent semantic targets. The student encoder is initialized from the same pretrained backbone but is updated during MEM training.
% Given an input image $I$, we randomly sample a binary mask $M$ that removes a subset of image patches. The teacher receives the masked image $M(I)$ and produces a semantic target representation:
Given an input image $I$, we sample a binary patch mask $m$ and obtain the masked image $\mathcal{M}(I; m)$. A frozen teacher visual encoder $\psi_I^{\mathrm{tea}}$ takes the masked image as input and produces the target representation:
\begin{equation}
    z_I^{\mathrm{tea}} = \psi_I^{\mathrm{tea}}(\mathcal{M}(I; m)) \in \mathbb{R}^{N_I \times D_I}.
\end{equation}
This target describes how the frozen teacher interprets the image when some visual regions are missing.

In parallel, the student visual encoder $\psi^{stu}_I$ receives the original image $I$ and produces patch-level visual tokens:
\begin{equation}
    f_I^{\mathrm{stu}} = \psi^{stu}_I(I) \in \mathbb{R}^{N_I \times D_I}.
\end{equation}
We then apply the same patch mask $m$ to select the student tokens corresponding to the visible regions:
\begin{equation}
    \widetilde{f}_I^{\mathrm{stu}} = \mathcal{P}_{m}(f_I^{\mathrm{stu}}),
\end{equation}
where $\mathcal{P}_{m}(\cdot)$ denotes the token selection operator induced by the visible patches under mask $m$. A lightweight prediction decoder $D(\cdot)$ takes the visible student tokens and predicts the teacher's masked-image representation:
\begin{equation}
    \hat{z}_I^{\mathrm{tea}} = D(\widetilde{f}_I^{\mathrm{stu}}).
\end{equation}
The MEM objective minimizes the distance between the predicted representation and the frozen teacher target:
\begin{equation}
    \mathcal{L}_{\mathrm{MEM}}
    =
    \left\|
    z_I^{\mathrm{tea}}
    -
    \hat{z}_I^{\mathrm{tea}}
    \right\|_2^2,
\end{equation}
This training process can be understood as a teacher-student learning problem. The teacher answers the question: ``How should the image be represented when this region is removed?'' The student is trained to predict this answer from its remaining unmasked patch tokens. Because different masks remove different regions, the student repeatedly observes how the semantic representation changes when each local region is absent. To solve this task well, the student must learn patch tokens that encode meaningful local information, such as which regions contribute evidence for opacity, cardiomegaly, pleural effusion, or other radiographic findings. Across many random masks, each patch is sometimes visible and sometimes removed. This encourages the student encoder to make each patch token locally informative, rather than relying only on a coarse global image representation.

% \hl{you keep talking about MAE. It is important to highlight what's lessons in MAE and how it relates to your problem. If there is some critical lesson in MAE reflecting the problem you are aiming, you should clarify and emphasize}

% MEM differs from standard MAE \hl{cite} in both its target and its purpose. MAE reconstructs missing pixels, which mainly teaches the model to recover low-level visual appearance. In contrast, MEM reconstructs the teacher's semantic representation of the masked image. Therefore, the student learns what semantic information is lost or preserved when a region is removed. This is more suitable for radiology report generation, where the key goal is not to reconstruct image texture, but to encode the local clinical evidence that supports the generated report.

After MEM training, the student visual encoder is used as the image encoder in our report generation framework. The decoder used only for MEM prediction is discarded. The resulting encoder provides visual tokens with stronger local semantics, making it easier for the language model to condition its generated report on actual image evidence.
% In our framework, MEM complements pathology-aware alignment: prototype alignment organizes image and text representations around shared pathology concepts, while MEM strengthens the local visual tokens that provide evidence for those concepts. Together, these two components encourage the model to generate reports that are both pathology-aware and visually faithful.

\subsection{Feature Extraction}
\label{sec:feat}

Given a chest X-ray image $I$ and its corresponding report $T$, we use a visual encoder $\psi_I$ (which is the student network in MEM), and a text encoder $\psi_T$ to extract image and report token representations:
\begin{equation}
\begin{aligned}
    f_I &= \psi_I(I) \in \mathbb{R}^{N_I \times D_I}, \
    f_T &= \psi_T(T) \in \mathbb{R}^{N_T \times D_T},
\end{aligned}
\end{equation}
where $N_I$ and $N_T$ denote the numbers of visual and textual tokens, respectively, and $D_I$ and $D_T$ are their feature dimensions.

The extracted visual and textual features are then projected into a shared latent space of dimension $D$ using modality-specific projection functions $\phi_I(\cdot)$ and $\phi_T(\cdot)$:
\begin{equation}
    \begin{aligned}
        F_I &= \phi_I(f_I) \in \mathbb{R}^{N_I \times D}, \
        F_T &= \phi_T(f_T) \in \mathbb{R}^{N_T \times D}.
    \end{aligned}
\end{equation}

We use the visual \texttt{[CLS]} token and textual \texttt{[CLS]} token as global modality-level descriptors:
\begin{equation}
    z_I = F_I^{\texttt{[CLS]}} \in \mathbb{R}^{D},
    \qquad
    z_T = F_T^{\texttt{[CLS]}} \in \mathbb{R}^{D}.
\end{equation}
These global image and report embeddings are then used for pathology-aware alignment.

\subsection{Pathology-Aware Alignment}
\label{sec:pathology_alignment}
As discussed above, directly aligning each image-report pair while treating all other samples as negatives may push apart cases that describe related or partially overlapping pathologies.
% This is undesirable in radiology, where different reports can share compatible pathology semantics even when they are not paired with the same image \hl{this sentence need to be rewritten}.
To address this issue, \ours{} introduces a pathology-aware alignment objective that aligns image and report representations through shared clinical concepts in a common vision--language embedding space. Instead of forcing pairwise separation, the alignment is guided by a set of pathology prototypes that serve as semantic anchors between radiographic evidence and textual findings.

\paragraph{Pathology Prototypes.}
We define a set of $K$ learnable pathology prototypes:
\begin{equation}
    P = [p_1, p_2, \dots, p_K]^\top \in \mathbb{R}^{K \times D},
\end{equation}
where each prototype $p_k$ represents a latent pathology concept in the shared embedding space. To provide clinically meaningful initialization, we initialize the prototypes by encoding pathology names from the CheXpert label set~\cite{irvin2019chexpert} using the text encoder $\psi_T$ and projection $\phi_T$ function. This encourages each prototype to start from a radiologically meaningful semantic concept while allowing it to be further optimized during training.

Given a mini-batch of $B$ image-report pairs, we collect the global image and report embeddings defined in Sec.~\ref{sec:feat} as $Z_I = [z_I^{(1)}, \ldots, z_I^{(B)}]^\top \in \mathbb{R}^{B \times D}$ and $Z_T = [z_T^{(1)}, \ldots, z_T^{(B)}]^\top \in \mathbb{R}^{B \times D}$. These embeddings are aligned to the shared pathology prototypes to build structured pathology-level correspondences across modalities.

\paragraph{Sinkhorn-Based Assignment.}
For each modality, let $Z \in {Z_I, Z_T}$ denote a batch of modality-level embeddings. We compute the similarity-based cost between sample embedding $z_i$ and pathology prototype $p_j$ as $C_{ij} = -z_i^\top p_j / \tau$, where $\tau$ is a temperature hyperparameter that controls the sharpness of the similarity distribution. We then estimate a soft assignment matrix by solving an entropy-regularized optimal transport problem \cite{peyre2019computational}:

% \begin{equation}
%     \min_{\mathcal{T}} ; \langle \mathcal{T}, C \rangle - \epsilon H(\mathcal{T}),
% \end{equation}
% \begin{equation}
%     \text{s.t.} \quad
%     \mathcal{T}\mathbf{1}_K = \frac{1}{B}\mathbf{1}_B,
%     \qquad
%     \mathcal{T}^\top\mathbf{1}_B = \frac{1}{K}\mathbf{1}_K,
% \end{equation}

\begin{equation}
\resizebox{\columnwidth}{!}{$
\displaystyle
\min_{\mathcal{T}} \; 
\langle \mathcal{T}, C \rangle - \epsilon H(\mathcal{T})
\quad \text{s.t.} \quad
\mathcal{T}\mathbf{1}_K = \frac{1}{B}\mathbf{1}_B,\;
\mathcal{T}^{\top}\mathbf{1}_B = \frac{1}{K}\mathbf{1}_K .
$}
\end{equation}

where $\langle\cdot,\cdot\rangle$ denotes the Frobenius inner product, $\mathcal{T}$ denotes the transport plan,
$H(\mathcal{T}) = -\sum_{ij}\mathcal{T}_{ij}\log\mathcal{T}_{ij}$ is the entropy regularizer, and $\epsilon$ controls the smoothness of the transport plan. This problem is solved efficiently using the Sinkhorn--Knopp algorithm~\cite{cuturi2013sinkhorn}. We denote the optimal transport plans obtained by the Sinkhorn algorithm for image-to-prototype and text-to-prototype matching as $\mathcal{T}_{I-P}$ and $\mathcal{T}_{T-P}$, respectively. These transport plans serve as soft assignment matrices between modality-specific representations and shared pathology prototypes.
\paragraph{Multi-Modal and Uni-Modal Consistency.}
The transport assignments indicate how strongly each image or report representation is associated with each pathology prototype. 
We first compute the prototype similarity distributions as $\mathcal{S}_{I-P}=\mathrm{softmax}(Z_I P^\top/\tau)$ and $\mathcal{S}_{T-P}=\mathrm{softmax}(Z_T P^\top/\tau)$.

We then impose a Multi-modal consistency loss that encourages image-based pathology assignments to match text-prototype similarities, and text-based pathology assignments to match image-prototype similarities:
\begin{equation}
    \mathcal{L}_{\mathrm{Multi}}
    =
    \mathcal{H}\big(\mathcal{T}_{I-P}, \mathcal{S}_{T-P}\big)
    +
    \mathcal{H}\big(\mathcal{T}_{T-P}, \mathcal{S}_{I-P}\big),
\end{equation}
where $\mathcal{H}(\cdot,\cdot)$ denotes the Cross-entropy loss.

To preserve modality-specific coherence, we further apply an Uni-modal consistency loss:
\begin{equation}
    \mathcal{L}_{\mathrm{Uni}}
    =
    \mathcal{H}\big(\mathcal{T}_{I-P}, \mathcal{S}_{I-P}\big)
    +
    \mathcal{H}\big(\mathcal{T}_{T-P}, \mathcal{S}_{T-P}\big).
\end{equation}
The final pathology-aware alignment objective is:
\begin{equation}
    \mathcal{L}_{\mathrm{align}} = \mathcal{L}_{\mathrm{Multi}} + \mathcal{L}_{\mathrm{Uni}}.
\end{equation}

By aligning image and report representations to shared pathology prototypes, \ours{} encourages semantically related cases to share common clinical concepts without pushing compatible samples apart. The OT assignment provides a balanced soft matching between mini-batch samples and pathology prototypes, preventing collapse to a few dominant disease concepts. Thus, $\mathcal{T}_{I-P}$ and $\mathcal{T}_{T-P}$ estimate how strongly each image or report is associated with different pathology concepts.
The prototype similarity distributions $\mathcal{S}_{I-P}$ and $\mathcal{S}_{T-P}$ represent the model's direct affinity predictions from embedding similarity. The uni-modal loss aligns these predictions with the OT assignments within each modality, while the multi-modal loss enforces consistency between the pathology distributions inferred from image and text. In this way, \ours{} aligns visual and textual features through shared pathology distributions rather than rigid instance-level matching, reducing the modality gap and improving the association between radiographic evidence and textual findings.

\subsection{Report Generation}
\label{sec:report_generation}

After learning pathology-aware visual--textual representations, we optimize the model for radiology report generation with a standard autoregressive objective. Given an input image $I$, we use the projected visual patch embeddings $F_I \in \mathbb{R}^{N_I \times D}$ from Sec.~\ref{sec:feat} as visual evidence for generation. In addition, when the Indication section is available, we encode it with the text encoder and projection function to obtain an indication feature $F_{\mathrm{Ind}} \in \mathbb{R}^{N_{\mathrm{Ind}} \times D}$. The report generator therefore conditions on both patch-level radiographic evidence and the clinical indication.

Specifically, the visual patch embeddings and indication features are fused through a lightweight transformer-based multimodal fusion module:
\begin{equation}
U = \mathrm{Fusion}(F_I, F_{\mathrm{Ind}}).
\end{equation}
The resulting multimodal representation $U$ is then provided to a DistilGPT2 decoder~\cite{sanh2019distilbert}, which generates the report tokens sequentially in an autoregressive manner:
\begin{equation}
    p(T \mid I, T_{\mathrm{Ind}})
    =
    \prod_{t=1}^{L}
    p(w_t \mid w_{<t}, U),
\end{equation}
where $T={w_t}_{t=1}^{L}$ denotes the ground-truth report. The generator is trained using the standard cross-entropy loss:
\begin{equation}
    \mathcal{L}_{\mathrm{gen}}
    =
    -\sum_{t=1}^{L}
    \log p(w_t \mid w_{<t}, U).
\end{equation}
This objective encourages the decoder to generate fluent reports conditioned on both aligned patch-level visual representations and the available clinical indication.

\section{Experiments}
\begin{table*}[t]
\centering
\setlength{\tabcolsep}{12pt}
\caption{Evaluation of descriptive accuracy for our proposed framework \ours{} versus existing state-of-the-art methods on the MIMIC-CXR, MIMIC-ABN, and IU X-ray datasets, using NLG metrics and RG. The best and the runner-up results are denoted in \textbf{bold} and \underline{underline}.}
\label{tab:main}
\resizebox{\textwidth}{!}{
\begin{tabular}{llrccccccc}
\toprule
\multirow{2}{*}{\textbf{Dataset}} 
& \multirow{2}{*}{\textbf{Method}} 
& \multirow{2}{*}{\textbf{Venue}} 
& \multicolumn{6}{c}{\textbf{NLG Metrics} $\uparrow$} 
& \multirow{2}{*}{\textbf{RG} $\uparrow$} \\
\cmidrule(lr){4-9}
& & & B-1 & B-2 & B-3 & B-4 & MTR & R-L \\
\midrule
% ---------------- M-CXR ----------------
\multirow{15}{*}{M-CXR}
& SA~\cite{yan2023style} & EMNLP'23 & - & 0.184 & - & - & - & - & 0.228 \\
& MET~\cite{wang2023metransformer} & CVPR'23 & 0.386 & 0.250 & 0.169 & 0.124 & 0.152 & 0.291 & - \\
& KIUT~\cite{huang-kiut} & CVPR'23 & 0.393 & 0.243 & 0.159 & 0.113 & 0.160 & 0.285 & - \\
& CoFE~\cite{cofe-eccv-24} & ECCV'24 & - & - & - & 0.125 & 0.176 & 0.304 & - \\
& MAN~\cite{shen2024automatic_aaai} & AAAI'24 & 0.396 & 0.244 & 0.162 & 0.115 & 0.151 & 0.274 & - \\
& B-LLM~\cite{aaai-liu2024bootstrapping-llm} & AAAI'24 & 0.402 & 0.262 & 0.180 & 0.128 & 0.175 & 0.291 & - \\
& DCG~\cite{liang2024divide-acmmm-24} & ACMMM'24 & 0.397 & 0.258 & 0.166 & 0.126 & 0.162 & 0.295 & - \\
& Med-LLM~\cite{liu2024in-context-acmmm} & ACMMM'24 & - & - & - & 0.128 & 0.161 & 0.289 & - \\
& SEI~\cite{sei} & MICCAI'24 & 0.382 & 0.247 & 0.177 & 0.135 & 0.158 & 0.299 & 0.249 \\
& FMVP~\cite{tmm_mulview_2024-fmvp} & TMIM'23 & 0.389 & 0.236 & 0.156 & 0.108 & 0.150 & 0.284 & - \\
& HERGen~\cite{2024-eccv-hergen} & ECCV'24 & 0.395 & 0.248 & 0.169 & 0.122 & 0.156 & 0.285 & - \\
& CXRMate~\cite{nicolson2023-longitudinal-multiview} & arXiv'23 & 0.361 & 0.223 & 0.150 & 0.108 & 0.159 & 0.263 & 0.238 \\
& LLaVa-Rad ~\cite{zambrano2025clinically} & Nature'25 & 0.365 & - & - & 0.144 & - & 0.300 & \underline{0.293} \\
& MLRG ~\cite{liu2025enhanced} & CVPR'25 & 0.411 & 0.277 & \underline{0.204} & \underline{0.158} & \underline{0.176} & \underline{0.320} & 0.291 \\
& DART ~\cite{park2025dart} & CVPR'25 & \underline{0.437} & \underline{0.279} & 0.191 & 0.137 & 0.175 & 0.310 & - \\
& \textbf{\ours} & - & \textbf{0.452} & \textbf{0.334} & \textbf{0.269} & \textbf{0.227} & \textbf{0.204} & \textbf{0.373} & \textbf{0.347} \\
\cmidrule(lr){2-10}
& $\Delta$ (abs.) $\uparrow$ & - & +0.015 & +0.055 & +0.065 & +0.069 & +0.028 & +0.053 & +0.056 \\
\midrule

% ---------------- M-ABN ----------------
\multirow{6}{*}{M-ABN}
& R2Gen~\cite{chen-etal-2020-generating} & EMNLP'20 & 0.253 & 0.144 & 0.092 & 0.063 & 0.106 & 0.229 & 0.179 \\
& CMN~\cite{chen-etal-2021-cross-modal} & ACL'21 & 0.256 & 0.147 & 0.095 & 0.066 & 0.110 & 0.230 & 0.183 \\
& SEI~\cite{sei} & MICCAI'24 & 0.267 & 0.157 & 0.104 & 0.073 & 0.114 & 0.231 & 0.191 \\
& MLRG ~\cite{liu2025enhanced} & CVPR'25 & \underline{0.332} & \underline{0.199} & \underline{0.132} & \underline{0.094} & \underline{0.136} & \underline{0.248} & \underline{0.219} \\
& \textbf{\ours} & - & \textbf{0.343} & \textbf{0.215} & \textbf{0.151} & \textbf{0.113} & \textbf{0.142} & \textbf{0.266} & \textbf{0.237} \\
\cmidrule(lr){2-10}
& $\Delta$ (abs.) $\uparrow$ & - & +0.011 & +0.016 & +0.019 & +0.019 & +0.006 & +0.018 & +0.018 \\
\midrule

% ---------------- IU-XRay ----------------
\multirow{11}{*}{IU-XRay}
& R2Gen~\cite{chen-etal-2020-generating} & EMNLP'20 & 0.470 & 0.304 & 0.219 & 0.165 & 0.187 & 0.371 & - \\
& CMN~\cite{chen-etal-2021-cross-modal} & ACL'21 & 0.475 & 0.309 & 0.222 & 0.170 & 0.191 & 0.375 & - \\
& PPKED~\cite{liu2021exploring} & CVPR'21 & 0.483 & 0.315 & 0.224 & 0.168 & 0.190 & 0.376 & - \\
& CMCL~\cite{liu2022competence} & ACL'21 & 0.473 & 0.305 & 0.217 & 0.162 & 0.186 & 0.378 & - \\
& MSAT~\cite{wang2022medical} & MICCAI'22 & 0.481 & 0.316 & 0.226 & 0.171 & 0.190 & 0.372 & - \\
& MET~\cite{wang2023metransformer} & CVPR'23 & 0.483 & 0.322 & 0.228 & 0.172 & 0.192 & 0.380 & - \\
& Med-LLM~\cite{liu2024in-context-acmmm} & ACMMM'24 & - & - & - & 0.168 & 0.209 & 0.381 & - \\
& MA~\cite{shen2024automatic} & AAAI'24 & \textbf{0.501} & 0.328 & 0.230 & 0.170 & \underline{0.213} & 0.386 & - \\
& B-LLM~\cite{liu2024bootstrapping} & AAAI'24 & \underline{0.499} & 0.323 & 0.238 & 0.184 & 0.208 & 0.390 & - \\
& DART ~\cite{park2025dart} & CVPR'25 & 0.486 & \underline{0.348} & \underline{0.265} & \underline{0.208} & 0.205 & \underline{0.411} & - \\
& \textbf{\ours} & - & 0.483 & \textbf{0.351} & \textbf{0.288} & \textbf{0.249} & \textbf{0.223} & \textbf{0.418} & \textbf{0.419} \\
\cmidrule(lr){2-10}
& $\Delta$ (abs.) $\uparrow$ & - & -0.018 & +0.003 & +0.023 & +0.041 & +0.010 & +0.007 & - \\
\bottomrule
\end{tabular}
}
\end{table*}
\begin{table}[t]
\centering
\setlength{\tabcolsep}{17pt}
\caption{Comparison of clinical efficacy metrics between the proposed method and state-of-the-art approaches, evaluated using F1 score, precision, and recall on the MIMIC-CXR dataset. The best and the runner-up are denoted in \textbf{bold} and \underline{underline}.}
\label{tab:ce}
\resizebox{\columnwidth}{!}{
\begin{tabular}{lccc}
\hline
\textbf{Model} & \textbf{F1} & \textbf{Precision} & \textbf{Recall} \\
\hline
R2Gen~\cite{chen-etal-2020-generating} & 0.276 & 0.333 & 0.273 \\
CMN~\cite{chen-etal-2021-cross-modal} & 0.278 & 0.334 & 0.275 \\
MET~\cite{wang2023metransformer} & 0.311 & 0.364 & 0.309 \\
Med-LLM~\cite{liu2024in-context-acmmm} & 0.395 & 0.412 & 0.373 \\
MA~\cite{shen2024automatic} & 0.389 & 0.411 & 0.398 \\
% LLaVa-Rad~\cite{zambrano2025clinically} & 0.530 & 0.411 & 0.398 \\
DART~\cite{park2025dart} & \underline{0.533} & 0.520 & \textbf{0.546} \\
MLRG~\cite{liu2025enhanced} & 0.505 & \underline{0.549} & 0.468 \\
\hline
\textbf{\ours (Ours)} & \textbf{0.545} & \textbf{0.597} & \underline{0.500} \\
\hline
\end{tabular}
}
\end{table}

\subsection{Experiment Settings}

\noindent\textbf{Datasets.} 
We evaluate our approach on three benchmark datasets: \textbf{MIMIC-CXR}~\cite{johnson2019mimic} is a large publicly available dataset of chest X-rays and associated radiology reports, containing around 377,110 images corresponding to 227,835 radiographic studies at the Beth Israel Deaconess Medical Center. \textbf{MIMIC-ABN}~\cite{ni2020learning} is a subset of MIMIC-CXR focusing specifically on abnormal findings with 72,000 images. \textbf{IU X-ray}~\cite{demner2015preparing} contains approximately 7,470 chest X-ray images paired with 3,955 radiology reports. We split following standard protocols with 70/10/20 for training, validation, and testing respectively.

\noindent\textbf{Metrics.} Following ~\cite{liu2025enhanced}, we evaluate SCOPE using natural language generation (NLG) and clinical efficacy (CE) metrics. NLG metrics, which measure the linguistic similarity between the generated and reference reports, include BLEU-n (B-n), METEOR (MTR), and ROUGE-L (R-L). 
For CE evaluation, we employ CheXpert~\cite{irvin2019chexpert} to label the generated reports with 14 clinical observations and compute micro-averaged Precision (P), Recall (R), and F1 score (F1) against the ground-truth annotations. Additionally, we adopt the F1 RadGraph (RG) metric~\cite{jain2021radgraph}, which quantifies the overlap of clinical entities and their relationships, providing a closer alignment with radiologists' assessments compared to traditional NLG and CE metrics.

\noindent\textbf{Implementation Details.} 
For the image encoder, we use a chest X-ray-pretrained DINOv2 \cite{perezrad} model, and for the text encoder, we adopt CXR-BERT \cite{boecking2022making}. We optimize the model using AdamW with a batch size of 8. The learning rate is set to $1\times10^{-5}$ for encoder fine-tuning and multimodal alignment, and $5\times10^{-5}$ for the report generation stage.

\noindent\textbf{Compared Baselines.}
We compare \ours{} with previous state-of-the-art radiology report generation methods, including R2Gen~\cite{chen-etal-2020-generating}, CMN~\cite{chen-etal-2021-cross-modal}, SA~\cite{yan2023style}, MET~\cite{wang2023metransformer}, KIUT~\cite{huang-kiut}, CoFE~\cite{cofe-eccv-24}, MAN~\cite{shen2024automatic_aaai}, B-LLM~\cite{aaai-liu2024bootstrapping-llm}, DCG~\cite{liang2024divide-acmmm-24}, Med-LLM~\cite{liu2024in-context-acmmm}, SEI~\cite{sei}, FMVP~\cite{tmm_mulview_2024-fmvp}, HERGen~\cite{2024-eccv-hergen}, CXRMate~\cite{nicolson2023-longitudinal-multiview}, PPKED~\cite{liu2021exploring}, CMCL~\cite{liu2022competence}, MSAT~\cite{wang2022medical}, MA~\cite{shen2024automatic}, LLaVa-Rad \cite{zambrano2025clinically}, DART~\cite{park2025dart}, and MLRG~\cite{liu2025enhanced}.
Since not all methods report results on every dataset, we include the available results from their original papers or the results reported in MLRG and DART. Due to limited computational resources and the unavailability of some official implementations or checkpoints, we do not reimplement missing baselines and leave unreported entries blank.
% Since not all methods report results on every dataset, we include the available results from their original papers. Due to limited computational resources and the unavailability of some official implementations or checkpoints, we do not reimplement missing baselines and leave unreported entries blank.

\subsection{Quantitative Results}
Table~\ref{tab:main} compares \ours{} with existing radiology report generation methods on MIMIC-CXR, MIMIC-ABN, and IU X-ray. Across the three datasets, \ours{} achieves strong and consistent performance on both NLG metrics and RG. On MIMIC-CXR, our method obtains the best results across all reported metrics, showing clear improvements over recent state-of-the-art methods. Similar trends can be observed on MIMIC-ABN, where \ours{} consistently outperforms previous approaches, indicating its effectiveness for reports containing abnormal and clinically specific findings.

On IU X-ray, \ours{} also achieves the best performance on most metrics, including BLEU-2, BLEU-3, BLEU-4, METEOR, ROUGE-L, and RG. Although some prior methods obtain slightly higher BLEU-1, \ours{} performs better on higher-order BLEU scores and other report-level metrics, suggesting that the generated reports better preserve longer textual structure and clinical content.

Table~\ref{tab:ce} further reports the clinical efficacy results on MIMIC-CXR. \ours{} achieves the best F1 score and precision among all compared methods, while maintaining competitive recall. These results suggest that the proposed framework improves not only descriptive report quality, but also the clinical accuracy of the generated findings.

\subsection{Qualitative Results}
\begin{figure}[t]
    \centering
    \includegraphics[width=\linewidth]{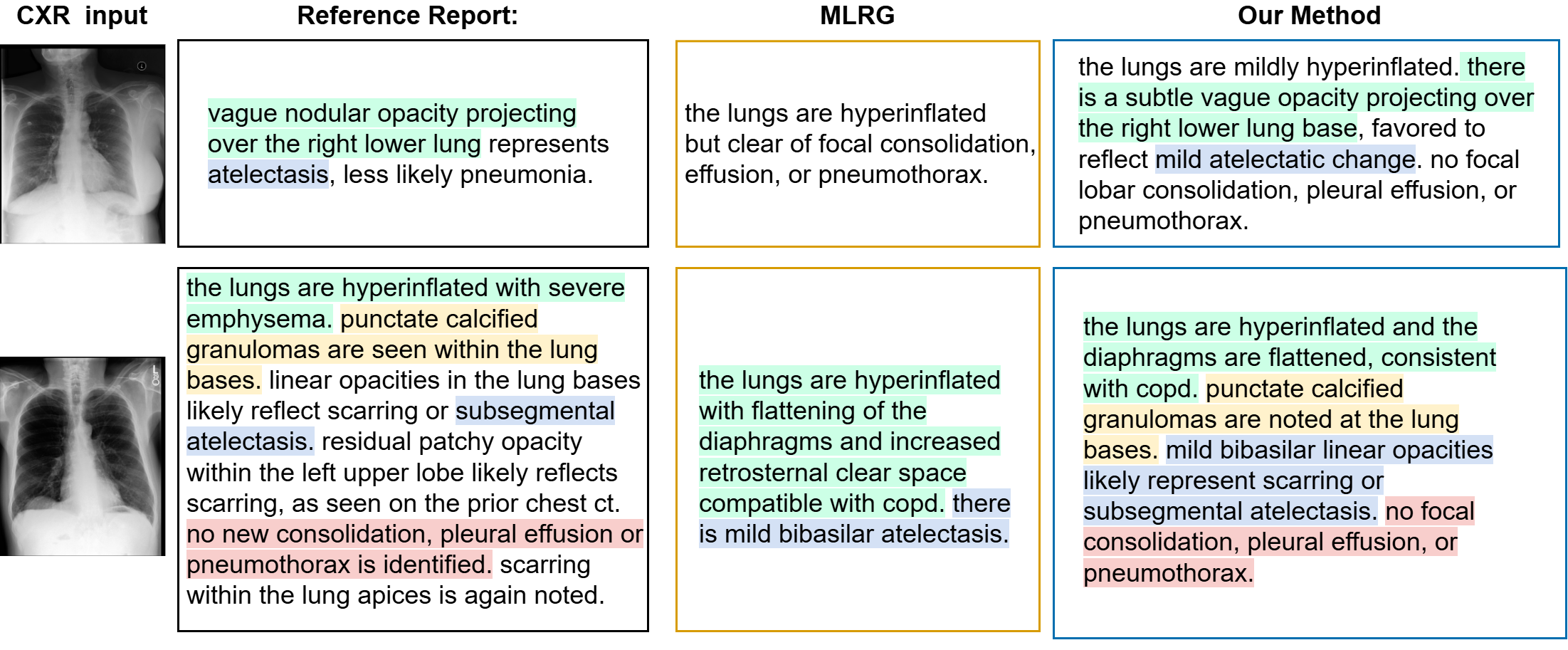}
    \caption{
    Qualitative comparison of generated reports. MLRG produces plausible but generic descriptions that miss subtle localized findings, while our method better captures radiographic evidence in the reference reports. Highlighted text denotes clinically relevant findings aligned with the reference.}
    \label{fig:quali}
\end{figure}

Figure~\ref{fig:quali} shows representative qualitative comparisons between MLRG and our method. MLRG often produces plausible but generic reports that capture dominant global patterns while missing subtle localized abnormalities. In the first example, the reference report identifies a vague right lower lung opacity favored to represent atelectasis, whereas MLRG describes the lungs as clear. Our method better preserves this localized finding by reporting a subtle right lower lung/base opacity with mild atelectatic change.

The second example further illustrates improved report completeness. Although MLRG partially captures the hyperinflation/COPD pattern, it omits several reference findings, including calcified granulomas and bibasilar linear opacities related to scarring or subsegmental atelectasis. Our method generates a more complete description that includes hyperinflation, calcified granulomas, bibasilar linear opacities, and the absence of focal consolidation, pleural effusion, or pneumothorax. These qualitative results indicate that our method better preserves clinically relevant radiographic evidence, especially subtle local findings and chronic abnormalities.

\subsection{Ablation Study}
\begin{table}[b]
\centering
\caption{
Ablation study on model components. PAA denotes Pathology-Aware Alignment, and MEM denotes Masked Evidence Modeling. A $\checkmark$ indicates that the component is used. For PAA, $\times$ means replacing the proposed pathology-aware alignment with a standard contrastive loss; for MEM, $\times$ means removing masked evidence modeling. NLG metrics include BLEU-4 (B-4), ROUGE-L (R-L), and METEOR (MTR), while clinical metrics include F1-RadGraph (RG) and F1.
}
\label{tab:ablation}
\setlength{\tabcolsep}{11pt}
\resizebox{\columnwidth}{!}{%
\begin{tabular}{c c c c c c c}
\toprule
\textbf{PAA} & \textbf{MEM} & \textbf{B-4} & \textbf{R-L} & \textbf{MTR} & \textbf{RG} & \textbf{F1} \\
\midrule
$\times$ & $\times$ & 0.166 & 0.331 & 0.175 & 0.296 & 0.473 \\
$\checkmark$ & $\times$ & 0.174 & 0.341 & 0.179 & 0.303 & 0.480 \\
$\times$ & $\checkmark$ & 0.186 & 0.345 & 0.184 & 0.311 & 0.489 \\
$\checkmark$ & $\checkmark$ & \textbf{0.227} & \textbf{0.373} & \textbf{0.204} & \textbf{0.347} & \textbf{0.545} \\
\bottomrule
\end{tabular}%
}
\end{table}

Table~\ref{tab:ablation} presents the ablation study of the main components in \ours{}. We evaluate the contribution of Pathology-Aware Alignment (PAA) and Masked Evidence Modeling (MEM). When PAA is not used, we replace it with a standard contrastive loss; when MEM is removed, the visual encoder is trained without masked evidence modeling.

Compared with the baseline setting without PAA and MEM, introducing either component improves the performance across both NLG and clinical metrics. PAA brings consistent gains by aligning visual and textual representations through shared pathology prototypes rather than relying only on direct contrastive matching. MEM also improves the results, suggesting that encouraging image tokens to encode local radiographic evidence benefits report generation.

The best performance is achieved when both PAA and MEM are used together. This shows that the two components are complementary: MEM enhances the visual representation by strengthening local evidence modeling, while PAA improves the semantic alignment between image and report representations. Their combination leads to the strongest results across BLEU-4, ROUGE-L, METEOR, RG, and F1.

\subsection{Evidence Sensitivity under Region Masking}
% \begin{table}[t]
% \centering
% \small
% \caption{Counterfactual region-masking analysis on RadGraph F1. Values in parentheses indicate the relative change compared with the unmasked image.}
% \label{tab:r-mask}
% \setlength{\tabcolsep}{20pt}
% \resizebox{\columnwidth}{!}{
% \begin{tabular}{llcc}
% \toprule
% \textbf{Model} & \textbf{Input Setting} & \textbf{RG} & \textbf{Change} \\
% \midrule
% \ours & Unmasked image & 0.330 & -- \\
% \ours & Left lung masked & 0.297 & -10.0\% \\
% \ours & Right lung masked & 0.302 & -8.5\% \\
% \ours & Both lungs masked & 0.246 & -25.5\% \\
% \ours & Background masked & 0.323 & -2.1\% \\
% \midrule
% MLRG & Unmasked image & 0.252 & -- \\
% MLRG & Left lung masked & 0.247 & -2.0\% \\
% MLRG & Right lung masked & 0.261 & +3.6\% \\
% MLRG & Both lungs masked & 0.225 & -10.7\% \\
% MLRG & Background masked & 0.246 & -2.4\% \\
% \bottomrule
% \end{tabular}
% }
% \end{table}

\begin{table}[t]
\centering
\small
\caption{Counterfactual region-masking analysis on RadGraph F1 and clinical F1.}
\label{tab:r-mask}
\setlength{\tabcolsep}{20pt}
\resizebox{\columnwidth}{!}{
\begin{tabular}{llcc}
\toprule
\textbf{Model} & \textbf{Input Setting} & \textbf{RG} & \textbf{F1} \\
\midrule
\ours & Unmasked image & 0.330 & 0.541 \\
\ours & Left lung masked & 0.297 & 0.532 \\
\ours & Right lung masked & 0.302 & 0.533 \\
\ours & Cardiac silhouette & 0.316 & 0.535 \\
\ours & Background masked & 0.323 & 0.539 \\
\midrule
MLRG & Unmasked image & 0.252 & 0.502 \\
MLRG & Left lung masked & 0.247 & 0.489 \\
MLRG & Right lung masked & 0.261 & 0.492 \\
MLRG & Cardiac silhouette & 0.225 & 0.489 \\
MLRG & Background masked & 0.246 & 0.499 \\
\bottomrule
\end{tabular}
}
\end{table}

We further evaluate whether generated reports respond appropriately to the removal of visual evidence. Following the region definitions in \cite{wu2021chest}, we mask clinically relevant anatomical regions and background regions, then evaluate the generated reports using RG and clinical F1. This analysis examines whether the model is affected more by the removal of meaningful radiographic evidence than by less informative background regions.
As shown in Table~\ref{tab:r-mask}, \ours{} consistently achieves better performance than MLRG across all masking settings. Although its performance decreases when anatomical regions are masked, this behavior is expected because important visual evidence has been removed. More importantly, \ours{} still maintains stronger report quality than MLRG under partial information loss, indicating better robustness when the input image is incomplete.
The results also show that \ours{} changes more clearly when clinically relevant regions are removed, while remaining relatively stable when the background is masked. This suggests that the model is sensitive to meaningful radiographic evidence rather than being strongly affected by uninformative perturbations. In contrast, MLRG shows weaker and less consistent changes under anatomical masking, suggesting greater reliance on learned report priors.

\section{Conclusion}

% In this work, we proposed \ours{}, a pathology-aware alignment framework for radiology report generation. Our method addresses the imperfect alignment between radiographic images and clinical reports, which often causes existing models to rely on pretrained language priors and generate clinically plausible but visually unsupported findings. To improve image-text alignment, \ours{} introduces shared pathology prototypes that organize visual and textual representations around clinically meaningful disease semantics, allowing semantically compatible cases to be aligned without treating them as negatives. In addition, we introduce Masked Evidence Modeling, a self-distillation objective that enhances the visual encoder's sensitivity to local radiographic evidence by encouraging patch-level representations to capture the semantic contribution of masked regions. Extensive experiments on MIMIC-CXR, MIMIC-ABN, and IU X-ray demonstrate that \ours{} achieves strong performance across both natural language generation and clinical efficacy metrics. These results suggest that explicitly modeling pathology-aware alignment and local visual evidence is an effective direction for generating more faithful and clinically informative radiology reports.
In this work, we proposed \ours{}, a pathology-aware alignment framework for radiology report generation. \ours{} addresses imperfect image-text alignment, which can cause models to rely on language priors and generate visually unsupported findings. Our method aligns visual and textual representations through shared pathology prototypes, allowing clinically compatible cases to be grouped by disease semantics without treating them as negatives. We further introduce Masked Evidence Modeling to improve the visual encoder's sensitivity to local radiographic evidence. Experiments on MIMIC-CXR, MIMIC-ABN, and IU X-ray show that \ours{} improves both language generation and clinical efficacy, supporting pathology-aware alignment as an effective direction for faithful radiology report generation.

\noindent\textbf{Limitations.}
Although \ours{} improves image-text alignment with pathology prototypes, the prototype set is initialized from predefined CheXpert labels, which may limit its ability to capture rare or fine-grained findings. Future work can explore adaptive pathology representations from clinical ontologies or large-scale medical foundation models.

% \noindent\textbf{Limitations.}
% Although \ours{} improves image-text alignment through pathology prototypes, the prototype set is initialized from predefined CheXpert labels. This limits its ability to capture rare, nuanced, or fine-grained pathological findings. Future work can explore richer and adaptive pathology representations derived from clinical ontologies or large-scale medical foundation models. 
% In addition, \ours{} requires an extra Masked Evidence Modeling stage to fine-tune the visual encoder, introducing additional training complexity compared with standard autoregressive report generation. More efficient evidence modeling strategies can help reduce this overhead.

% \noindent\textbf{Limitations.}
% Although \ours{} leverages pathology prototypes to improve image-text alignment, these prototypes are initialized from a predefined CheXpert-based pathology set. Consequently, the limited diversity of the prototype vocabulary may restrict the model's ability to capture rare, nuanced, or fine-grained pathological variations. Incorporating more comprehensive and adaptive pathology representations, such as prototypes derived from richer clinical ontologies or large-scale medical foundation models, could further enhance our framework. Another limitation is that \ours{} requires an additional visual encoder fine-tuning process through Masked Evidence Modeling, which may introduce extra training complexity compared with standard autoregressive report generation models. Future work could explore more efficient evidence modeling strategies to reduce this overhead.

\newpage
{
    \small
    \bibliographystyle{ieeenat_fullname}
    \bibliography{main}
}

\end{document}